\documentclass[12pt]{article}
\usepackage{amsfonts}
\usepackage{latexsym}
\usepackage{amsmath,amssymb}
\usepackage{verbatim}
\usepackage{setspace}
\usepackage{xcolor}
\usepackage{cite}

\usepackage[textheight=9in, textwidth=6.5in, letterpaper]{geometry}
\def\half{{1\over 2}}
\numberwithin{equation}{section}

\def\e{{\epsilon}}

 \def\p{\partial}
 \def\bz{{\bar z}}
 
\def\0{{(0)}}
\def\1{{(1)}}
\def\2{{(2)}}

\def\n{\nabla}

\def\<{\langle }
\def\>{\rangle }
\def\[{\left[}
\def\]{\right]}

\def\z{{\zeta}}

\newcommand{\bea}{\begin{eqnarray}}
\newcommand{\eea}{\end{eqnarray}}
\newcommand{\be}{\begin{equation}}
\newcommand{\ee}{\end{equation}}
\newcommand{\ba}{\begin{align}}
\newcommand{\ea}{\end{align}}

\renewcommand{\O}{\Omega}

\renewcommand{\epsilon}{\varepsilon}

   \makeatletter
  \let\over=\@@over \let\overwithdelims=\@@overwithdelims
  \let\atop=\@@atop \let\atopwithdelims=\@@atopwithdelims
  \let\above=\@@above \let\abovewithdelims=\@@abovewithdelims
\renewcommand\section{\@startsection {section}{1}{\z@}%
                                   {-3.5ex \@plus -1ex \@minus -.2ex}
                                   {2.3ex \@plus.2ex}%
                                   {\normalfont\large\bfseries}}

\renewcommand\subsection{\@startsection{subsection}{2}{\z@}%
                                     {-3.25ex\@plus -1ex \@minus -.2ex}%
                                     {1.5ex \@plus .2ex}%
                                     {\normalfont\bfseries}}

\begin{document}

\vspace{4.0cm}

\pagestyle{empty}
\pagestyle{plain}

\def\gzz{\gamma_{z\bz}}
\def\vx{{\vec x}}
\def\p{\partial}
\def\po{$\cal P_O$}
\def\cN{{\cal N}_\rho^2 }
\def\N{${\cal N}_\rho^2 ~~$}
\def\G{\Gamma}
\def\a{{\alpha}}
\def\b{{\beta}}
\def\g{\gamma}
\def\ch{{\cal H}}
\def\t{\theta }
\def\O{\Omega}
\def\cq{{\cal Q}}
\def\o{\omega}
\def\slsl{$SL(2,R)_L\times SL(2,R)_R$}
\def\vlvl{${\rm Vir_L}\times{\rm Vir_R}$}
\def\sb{\Sigma_{\rm bif}}
\pagenumbering{arabic}

\vskip1cm
\begin{center}
{~\\[140pt]{ \LARGE {\textsc{Kerr-Newman Black Hole Entropy and Soft Hair }}}\\[-20pt]}
\vskip2cm

\vspace{0.8cm}
Sasha Haco$^{* }$, ~ Malcolm J. Perry$^{*}$ and Andrew Strominger$^{\dagger}$

\vspace{1cm}

\begin{abstract}
   Recently a set of diffeomorphisms were found which act nontrivially on the Kerr horizon and form a left-right pair of Virasoro algebras. Using  the boundary formula for the associated central charge and assuming applicability of the Cardy formula to a putative dual CFT,  the Bekenstein-Hawking  entropy for the Kerr black hole was reproduced. In this paper, the addition of charge to the black hole is shown to require a minimal modification to this construction which then  reproduces the Bekenstein-Hawking entropy for the Kerr-Newman black hole.
   
 \end{abstract}
\vspace{0.5cm}

\vspace{2.0cm}

\end{center}


\noindent{*DAMTP, Cambridge University, Centre for Mathematical Sciences, Wilberforce Road, Cambridge CB3 0WA, UK}

\noindent{$\dagger$Center for the Fundamental Laws of Nature, Harvard University, Cambridge, MA USA}

\vskip2cm
\tableofcontents

\newpage 
\section{Introduction}
Diffeomorphisms (diffeos) in general relativity can act non-trivially on both the classical phase space and the physical quantum states whenever the spacetime has a boundary. The structure of such diffeos is exceedingly rich. Explorations of this subject began over half a century ago with  \cite{Bondi:1962px, Sachs:1962wk} (recent adaptations to the black hole context include  \cite{Hawking:2016msc,Hawking:2016sgy,Flanagan:2015pxa, Averin:2016ybl, Compere:2016jwb, Sheikh-Jabbari:2016lzm, Baxter:2016nml, Compere:2016gwf, Mao:2016pwq, Averin:2016hhm, Cardoso:2016ryw, Mirbabayi:2016axw, Grumiller:2016kcp, Donnay:2016ejv, Gabai:2016kuf, Gomez:2016hxz, He:2016yhv, Tamburini:2017dig, Ammon:2017vwt, Zhang:2017geq, Bousso:2017dny,  Hotta:2017yzk, Mishra:2017zan, Gomez:2017ioy, Grumiller:2017otl, Chatterjee:2017zeb, Chu:2018tzu, Kirklin:2018wvq, Cvetkovic:2018dmq, Grumiller:2018scv, Chandrasekaran:2018aop, Averin:2018owq, Choi:2018oel, Donnay:2018ckb, Compere:2016hzt, Donnay:2015abr}) but it remains only partially understood. The subject  has been put on firmer footing with the discovery \cite{Strominger:2014pwa, Strominger:2013jfa, He:2014laa} that in some cases the diffeos are equivalent to well understood soft graviton insertions in quantum field theory \cite{Weinberg:1965nx}. 

In string theory, such diffeos which act nontrivially at the horizon have been used to count black hole microstates and microscopically reproduce the macroscopic  area law \cite{Strominger:1996sh}.  In \cite{Hawking:2016msc, Hawking:2016sgy}, it was shown that nontrivial diffeos act on the horizon of a generic 4D Kerr  black hole and create distinguishing features referred to as soft hair. In  \cite{Haco:2018ske}, following the earlier work  \cite{Castro:2010fd} on hidden conformal symmetry, a set of diffeos of a spin $J$ Kerr were found with the properties (i) the   Lie bracket algebra acts as  a  left-right Virasoro pair on the horizon, 
(ii) the corresponding Wald-Iyer-Zoupas charges \cite{Crnkovic:1986ex,Zuckerman:1989cx,Lee:1990nz,Brown:1992br,Iyer:1995kg,wz,Barnich:2001jy} with a judicious choice of counterterm, have central terms  with $c_L=c_R=12J$. Assuming the existence of  a unitary Hilbert space (including horizon edge states) transforming under these Virasoros, along with the Cardy formula, the area law then follows. 

A Kerr black hole is characterized by two quantities, the mass $M$ and spin $J$. It  therefore cannot correspond to a general thermal state in a (parity symmetric) 2D CFT, which would be described by three parameters: the central charge $c_R=c_L$ and left and right temperatures $T_R$ and $T_L$. What happens is that the Kerr black hole corresponds to thermal states with temperatures constrained by the relation  $T_L^2=T_R^2+1$.

In this paper we add charge $Q$ and generalize our previous work \cite{Haco:2018ske} to the Kerr-Newman black hole. We heavily rely on Wang and Liu \cite{Wang:2010qv}, in which the notion of hidden conformal symmetry was generalized from Kerr to Kerr-Newman black holes. Interestingly the addition of a charge parameter relaxes the constraint between $T_R$ and $T_L$ which become independent variables.  That is the main difference between Kerr and Kerr-Newman for our purposes. 
The conformal-coordinate expression for the near horizon geometry in terms of $T_R$ and $T_L$ given in \cite{Haco:2018ske} is unchanged,  and the analysis proceeds in a nearly identical fashion to that of the neutral Kerr. In particular, the macroscopic area law for Kerr-Newman also follows from the assumption of a Cardy formula governing the black hole microstates.

\section{Hidden conformal symmetry}
Hidden conformal symmetry of the 4D Kerr black hole was identified \cite{Castro:2010fd} by examining the scalar wave equation of soft modes in the near-horizon region. This is a region of the phase space ($not$ of the geometry) in which
\be \omega M \ll 1,  \: r \omega \ll 1 .\ee In this region the solutions are  hypergeometric functions of $r$ which  fall into representations of an $SL(2,R)$ conformal symmetry. The scalar wave equation can be  written as the Casimir operator of a set of vector fields either with an $SL(2,R)_R$ or $SL(2,R)_L$ Lie bracket algebra. These `hidden' symmetries are broken by the  the azimuthal angular identification. This allows for a canonical identification of left and right temperatures $T_L$ and $T_R$ (see formulae below). 

Another way to glimpse  the symmetry is from the near region contribution to the soft scalar absorption cross section 
\bea\label{abs} {\cal P}_{abs}\sim T_L^{2h_L-1}T_R^{2h_R-1}\sinh({\o_L  \over 2T_L}+{\o_R  \over 2T_R} )  \big|\G(h_L+i{\o_L \over 2\pi T_L})\big|^2  \big|\G(h_R+i{\o_R  \over 2\pi T_R})\big|^2, \eea
where for a scalar $h_L=h_R=\ell$ with $\ell$ the angular momentum and $\omega_{L,R}$ are thermodynamically conjugate to $T_{L,R}$.
This  precisely matches that of the absorption cross section of an energy $(\omega_L, \omega_R)$ excitation of a 2D CFT at temperatures $(T_L, T_R)$. 

This structure is perhaps  a hint that a  hidden
conformal symmetry, which acts not just on the geometry but on the phase space,  is relevant to the structure of all black holes, not just extremal ones. If this is the case, and the CFT has central charges $c_L = c_R = 12J$ and obeys  the Cardy formula, one can reproduce the entropy of the black hole \cite{Castro:2010fd}. It is of interest to try to push this speculative idea further in various directions and make it more explicit.\footnote{One interesting direction would be to investigate the crossover Kac-Moody structure discussed in \cite{Detournay:2012pc}.}

This hidden conformal symmetry was subsequently shown to also be present in the case of 4D Kerr-Newman black holes \cite{Wang:2010qv, Chen:2010xu, Chen:2010zwa, Chen:2010ywa}.  Here, the near-horizon neutral scalar wave equation for soft modes  exhibits this same behavior, with the additional constraint,
\be \omega Q \ll 1
.\ee
The Kerr-Newman entropy is 
\be S_{BH} =  \pi (r_+^2 + a^2) =  \pi (2 M r_+ - Q^2).\ee 
where the outer event horizon $r_+$ and the inner Cauchy horizon $r_-$ are defined by 
\be r_\pm = M \pm \sqrt{M^2 - a^2 - Q^2},\ee and $a={J \over M}.$
The first law is 
\be T_H \delta S_{BH} = \delta M - \Omega \delta J - \Phi \delta Q , \ee where $T_H$ is the Hawking temperature, given by  \be T_H={r_+-r_-\over 4\pi (r_+^2 + a^2)}, \ee  the angular velocity of the horizon $\Omega$ is \be \Omega={a \over r_+^2 + a^2} \ee and $\Phi$ is the electric potential of the Kerr-Newman black hole, \be \Phi = {Q r_+ \over r_+^2 + a^2} . \ee The first law may also be written \be \delta S_{BH} = {\delta E_L \over T_L} + {\delta E_R \over T_R} , \ee 
where 
\bea \delta E_L &=& {2 M^2 - Q^2 \over a } \delta M + {Q ( Q^2 - 2 M^2) \over 2 J} \delta Q \cr \delta E_R &=& {2 M^2 - Q^2 \over a } \delta M - \delta J -  {Q M \over a} \delta Q,  \eea 
and  the left and right temperatures are defined by,
\bea \label{tmp} T_L &=& {r_+ + r_- \over 4 \pi a} - {Q^2 \over 4 \pi M a}, \\ T_R &=& {r_+ - r_- \over 4 \pi a} \eea 
For a neutral scalar 
\be \delta M = \omega  , \,  \delta J = m , \, \delta Q = 0, \ee  
 with $\o$ and $m$ being the soft mode scalar  energy and angular momentum operators. The Frolov-Thorne vacuum density matrix for such a scalar is (up to normalization) 
\be \rho_{FT}=e^{-{\o \over T_H}+{\Omega m \over T_H}+ {\Phi e \over T_H}} =e^{-{\delta E_R \over T_R}-{\delta E_L \over T_L}}.\ee
The left/right energies are then given in terms of the left/right-moving frequencies by \cite{Chen:2010xu},
\bea \label{ewt} \delta E_L &=& \omega_L = {2 M^2 - Q^2 \over a} \omega, \cr \delta E_R &=& \omega_R = {2 M^2 - Q^2 \over a} \omega - m ,
\eea 
 with $(\omega,m)$ the soft mode energy and axial component of angular momentum. 
 Using these modified definitions for Kerr-Newman, one then finds that the soft scalar 
absorption \eqref{abs}, originally derived for Kerr, remains valid. 

In \cite{Haco:2018ske}, this numerological discussion  was brought into sharper focus for the case of the Kerr black hole by providing a set of $Vir_{\,L} \otimes Vir_{\,R}$ vector fields which generate the full symmetry. These vector fields were used to compute the central charges in the covariant phase space formalism, yielding $c_L = c_R = 12J$. Here, the same argument is followed, with minor modifications,  for  the case of the Kerr-Newman black hole.

\section {Conformal coordinates} 
The Kerr-Newman metric in Boyer-Lindquist coordinates is 
\bea ds^2&=&-\big({\Delta - a^2 \sin^2\theta \over \rho^2}\big)dt^2+\big({(r^2+a^2)^2 - \Delta a^2 \sin^2\theta \over \rho^2}\big)\sin^2\theta d\phi^2 \cr &&~~~~~~~~~~~~~ - \big({2 a^2 \sin^2\theta (r^2 + a^2 - \Delta) \over \rho^2}\big) d\phi dt
 +{\rho^2 \over \Delta}dr^2+\rho^2 d\theta^2,\eea
where \bea
~~~~~~\rho^2=r^2+a^2\cos^2\theta, ~~~~~~\Delta=r^2+a^2+Q^2-2Mr.\eea
The gauge field $A$ is
\be A = - {Q r \over \rho^2} (dt - a \sin^2 \t d\phi). \ee 
Conformal coordinates  are \cite{Castro:2010fd}
\bea
 w^+&=& \sqrt{r-r_+ \over r-r_-}e^{{2\pi T_R}\phi},\cr
 w^-&=& \sqrt{r-r_+ \over r-r_-}e^{{2\pi T_L}\phi-{t \over 2M}},\cr
y&=& \sqrt{r_+ -r_-\over r-r_-}e^{{\pi(T_R+T_L)}\phi-{t \over 4M}}.\eea These are the same as defined in \cite{Castro:2010fd}, but note the different $Q$-dependent definitions of the temperatures \eqref{tmp} are used here. As in \cite{Haco:2018ske}, it can be shown that the past horizon is at $w^+=0$, the future horizon at $w^-=0$ and the bifurcation surface $\sb$ is at $w^\pm=0$. Under azimuthal identification $\phi \to \phi+2\pi$, the coordinates have the periodicities,
\be\label{idn} w^+ \sim e^{4\pi^2 T_R}w^+, ~~ w^- \sim e^{4\pi^2 T_L}w^-, ~~
y\sim e^{2\pi^2 (T_R+T_L)}y.\ee
Writing the Kerr-Newman metric in conformal coordinates, to leading and subleading  order around the bifurcation surface, we get
  \bea \label{metric} \begin{split} ds^2&= {4 \rho_+^2 \over y^2} d w^+ dw^-
+  {16 J^2 \sin^2\theta \over y^2 \rho_+^2} dy^2 +\rho_+^2 d\theta^2 \\[6pt]  &
- {2w^+ (8\pi J)^2 T_R(T_R+T_L)  \over y^3 \rho_+^2} dw^- dy \\[6pt]  &
+ {8 w^- \over y^3 \rho_+^2} \big(- (4\pi J)^2T_L(T_R+T_L) + (4 J^2 + 4\pi J a^2 (T_R+T_L)  + a^2 \rho_+^2) \sin^2\theta \big) dw^+ dy \\[6pt] &
+ \cdots, \end{split} \eea  
where corrections are at least second order in $(w^+,w^-)$. This metric takes precisely the same form as the Kerr black hole (as in \cite{Haco:2018ske}), but again with different definitions of $T_L, T_R$ and hence of $w^+, w^-, y$.

\section{Conformal vector fields}
Consider the same set of vector fields presented in \cite{Haco:2018ske} 
 \bea \zeta_n = \e_n \p_++\half\p_+\e_n y\p_y, & \e_n={2 \pi T_R}(w^+)^{1+{in \over 2 \pi T_R}}, \cr \bar \zeta_n = \bar \e_n \p_-+\half\p_-\bar \e_ny\p_y, &
\bar\e_n={2 \pi T_L}(w^-)^{1+{in \over 2 \pi T_L}},\eea
 so that $\z$ and $\bar \zeta $ are invariant under $2\pi$ azimuthal rotations \eqref{idn}. 
 These vector fields commute with one another and each obey a centreless Virasoro algebra,
 \be [\z_m,\z_n]=i(n-m)\z_{n+m},\ee and similarly for $\bar{\z}$.
 Their zero modes are 
 \bea \label{rtj} \zeta_0&=&2\pi T_R( w^+\p_++\half y\p_y )=\p_\phi+{2M^2 - Q^2 \over a} \p_t=-i\omega_R, \cr 
 \bar \z_0 &=& 2\pi T_L( w^-\p_-+\half y\p_y )=-{2M^2 - Q^2\over a}\p_t=i\o_L 
 \eea
 where the right and left moving energies $\o_R, \o_L$ are defined in \eqref{ewt}.

\section{Covariant charges}
Here we employ the construction of covariant charges as developed in, for example,  \cite{Crnkovic:1986ex,Zuckerman:1989cx,Lee:1990nz,Brown:1992br,Iyer:1995kg,wz,Barnich:2001jy}. 
The construction of covariant phase space charges begins with the Lagrangian,
\be \mathcal{L} = {\sqrt{-g} \over 16 \pi} ( R - F_{ab} F^{ab}), \ee where $R$ is the Ricci scalar and $F = dA $ is the electromagnetic field strength. 
Upon varying the field strength, $\delta A_a$ and the metric, $\delta g^{ab} = h^{ab}$ in
the Lagrangian, we get the presymplectic potential three-form, ${\bf{\Theta}} = * \t$, where
\be \theta[h, g, A, F] =  (\t_G[h,g] + \t_E[A,F])_a dx^a \ee 
where $\t_G$ is the gravitational part of the presymplectic potential,
\be \t_G^a [h,g] = - {1 \over 16 \pi}(\n_bh^{ab} - \n^a h) \ee 
and $\t_E$ is the part arising from the electromagnetic piece of the Lagrangian,
\be \t_E^a[A,F] = -  {1 \over 4 \pi} F^{ab} \delta A_b .\ee 
When the metric variation is due to a diffeo $\z$ and $A$ has a gauge transformation $\lambda$, i.e. 
\bea 
\delta g^{ab} &=& h^{ab} = \mathcal{L}_\z g^{ab}, \cr \delta A_a &=& (\mathcal{L}_\z A)_a + \n_a \lambda , \eea 
then provided the background field equations are satisfied, the Noether charge density two-form, ${\bf{Q_N}} = * Q_N$, is defined
\be d {\bf{Q}}_N = {\bf{\Theta}} - \iota_{\z} L, \ee
where $L = * \mathcal{L}$ and the gauge transformation $\lambda$ will be fixed below.
Thus,
\bea (Q_N^{ ab})_G &=& - {1 \over 16 \pi} (\n^a \z^b - \n^b \z^a) \cr (Q_N^{ ab})_E &=&  - {1 \over 4 \pi} F^{ab} (A_c \z^c + \lambda)
\eea 
The general form for the linearized charge associated to a diffeo $\z$ on a surface $\Sigma$ with boundary $\p\Sigma$  is \cite{wz} 
\be \delta \cq= {1 \over 16\pi} \int_{\p\Sigma} {\bf{k}},\ee
where the symplectic 2-form charge integrand can be found from
\be {\bf{k}} = \delta {\bf{Q}}_N - \iota_\z {\bf{\Theta}}. \ee 
With ${\bf{k}} = * (k_G + k_E)$, the gravitational part is
\be \begin{split} k^{ab}_G = {1 \over 16 \pi}\big[ \frac{1}{2}\n^a\zeta^bh
+\n^ah^c{}^b \zeta_c
+\n_c\zeta^a\ h^{bc} 
+\n_ch^{ac}\ \zeta^b
-\n^ah\ \zeta^b\big] - (a \leftrightarrow b) . \end{split} \ee
This makes up the Iyer-Wald charge.
The part due to the electromagnetic field strength is \cite{Hajian:2015xlp},
\be k_E^{ab} = - {1 \over 8 \pi} \big[ (\delta F^{ab} - {h \over 2} F^{ab} + 2 F^{ad} h_d^b)(A_c \z^c + \lambda) + F^{ab} \delta A_c \z^c  - 2 F^{ac} \delta A_c \z^b\big] - ( a \leftrightarrow b) . \ee 

Let us first consider the electromagnetic part of the charge.  
At this stage we  fix the gauge freedom  $\lambda$ with the judicious choice 
\be \lambda = - A_c \z^c \ee 
so that all but the last two terms in $k^{ab}_E$ vanish.
On the future horizon, where $w^- = 0$, the components that contribute to the integral are $k_E^{-y}$ and $k_E^{-+}$. 
Since $\z^- = 0$, all terms in the calculation involve the components of the electromagnetic field strength where one index is $w^-$, i.e. either $F^{-y}$, $F^{-+}$ or the $F^{-\t}$ components. It is straightforward to compute these components in conformal coordinates and one finds that they are either zero or linear in $w^-$. Since the metric contains no poles in $w^-$, these components will vanish on the future horizon. This means that there is no contribution to the charge integral from $k_E$.

Therefore the entire contribution to the charge arises from the gravitational part. 

As explained in \cite{Haco:2018ske}, the construction of these covariant charges involves many subtleties and ambiguities. The Iyer-Wald charge associated to the diffeomorphisms $\z_n$ is built from the symplectic form using  the metric variation is due to this diffeomorphism. This formalism alone is found to be inadequate for the construction of well-defined charges, as integrability and associativity may be violated. This necessitates the addition of certain counterterms, as set out by Wald and Zoupas, and discussed, for example in \cite{Lee:1990nz,Iyer:1995kg,wz,Barnich:2001jy,Chandrasekaran:2018aop}. It is equally possible that counterterms may arise for the electromagnetic part of the charge, due to similar ambiguities. However, since the electromagnetic part as defined gives zero contribution to the charge, one can not use arguments from integrability or associativity to motivate such an addition. 

The general form for the linearized charge associated to a diffeo $\z$ on a surface $\Sigma$ with boundary $\p\Sigma$  is \cite{wz} \be \label{charge}\delta \cq=\delta \cq_{IW}  +\delta \cq_X,\ee
where the Iyer-Wald charge is given above from the gravitational part of the presymplectic form,
\be \delta \cq_{IW}(\z,h;g)={1 \over 16\pi }\int_{\p\Sigma}*k_{G},\ee
and the Wald-Zoupas counterterm is 
\be \delta \cq_X={1 \over 16\pi }\int_{\p\Sigma}\iota_\z (* X),\ee
where $X$  is a spacetime one-form constructed from the geometry and linear in $h$.

The choice of counterterm for the gravitational case is not well-understood. It is not a priori fully determined by the considerations of \cite{Iyer:1995kg,wz}, where its precise form is left as an ambiguity. It was however shown \cite{Haco:2018ske} in the present context that, without a counterterm, there is an obstruction to defining integrable charges which canonically generate the symmetry. Moreover this obstruction can be eliminated by the counterterm choice  with \cite{Haco:2018ske} 
\be \label{counter}  X = 2dx^ah_a^{~b} \Omega_b, \ee where $\Omega_a$ is the H\'a\'{\j}i\v cek one-form,
\be \Omega_a = q_a^c n^b \n_c l_b, \ee 
a measure of the rotational velocity of the horizon. 
The null  vectors $\ell^a$ and $n^a$  are both normal  to $\sb $ and normalized such that $\ell \cdot n=-1$. $\ell$ ($n$) is taken to be normal to the future (past) horizon. $q_{ab}=g_{ab}+\ell_an_b+n_a\ell_b$ is the induced metric on $\sb$. It has not been shown that this choice is unique in eliminating the obstruction or that the charges so obtained do indeed canonically generate the symmetry. We nevertheless continue to use this formalism to construct the charges for the Kerr-Newman black hole.

Since the Kerr-Newman metric \eqref{metric} is identical to the corresponding metric in Kerr (albeit with different definitions of the temperatures), the calculations of the charges with respect to the same vector fields will be identical.  The resulting central terms are therefore (see \cite{Haco:2018ske} for a full derivation of the charges),

\be c_L = c_R = 12J \ee 

   \section{The area law}
Using  $c_L=c_R=12J$ as given above, the temperature formulae \eqref{tmp} and the Cardy formula
\be S_{Cardy}={\pi^2 \over 3}(c_LT_L+c_RT_R),\ee
yields the Bekenstein-Hawking  area-entropy law for generic Kerr
\be S_{BH}=S_{Cardy}=\pi (2 Mr_+ - Q^2) = {Area \over 4 }.\ee

\section{Acknowledgements}
This work was supported in part by DOE DE-sc0007870, the UK STFC and the John Templeton Foundation via the Black Hole Initiative.


\end{document}